\documentclass[12pt,preprint]{aastex}

\slugcomment{Draft version June 9, 2006.}

\shorttitle{Fluorine abundances in cool EHes}
\shortauthors{Pandey et al.}

\begin{document}

\title{Discovery of Fluorine in Cool Extreme Helium Stars}

\author{Gajendra Pandey}
\affil{Indian Institute of Astrophysics;
Bangalore,  560034 India}
\email{pandey@iiap.res.in}

\begin{abstract}
Neutral fluorine (F\,{\sc i}) lines are identified in the optical
spectra of cool EHe stars. These are the first
identification of F\,{\sc i} lines in a star's spectrum, and
provide the first measurement of fluorine
abundances in EHe stars. The results show that fluorine
is overabundant in EHe stars. The overabundance of
fluorine provides evidence for the synthesis of fluorine in
these stars, that is discussed in the light of asymptotic giant branch (AGB)
evolution, and the expectation from accretion of an He white dwarf by a C-O
white dwarf.
\end{abstract}

\clearpage
\keywords{stars: abundances --
stars: chemically peculiar -- stars: evolution}

\section{Introduction}

The solar system, for a long time, has been the only source of information about the 
fluorine abundance in the Galaxy \citep{hall69}. The astrophysical origin of solar 
system's fluorine is not yet identified from the known 
theories of stellar nucleosynthesis. The major problem with fluorine
production is that, the element has only one stable, yet rather fragile,
isotope, $^{19}$F. In stellar interiors it is readily destroyed by hydrogen
via $^{19}$F(p,$\alpha$)$^{16}$O and helium via $^{19}$F($\alpha$,p)$^{22}$Ne.
The high F abundances which were measured, using infrared HF vibration-rotation transitions,
in the asymptotic giant branch (AGB) stars from the \citet{jori92} sample, provided clues 
for fluorine production in AGB stars.
The \citet{cunha2003} study measured F abundances in different populations:
in galactic red giants (a subset of Jorissen et al. sample), in red giants in the Large
Magellanic Cloud (LMC), as well as in a few stars in the globular cluster Omega Centauri.
Concerning $\omega$\,Cen, the F abundances found in that study were quite low. One
possible interpretation was that the AGB stars in $\omega$\,Cen did not contribute 
significantly to the fluorine. However, this conclusion assumes that there has been
effective AGB pollution in $\omega$\,Cen.

Apart from the third dredge-up of AGB stars, the other two sources of
fluorine are Type\,II supernovae (SNe) explosions, and stellar winds
from Wolf-Rayet (W-R) stars. \citet{woos95} showed that in 
Type\,II SNe, $^{19}$F can be created by spallation of $^{20}$Ne by 
neutrinos. In massive or W-R stars, $^{19}$F is probably produced during
He-burning phase and, before it gets annihilated, is ejected into space
by strong stellar winds \citep{may2000}. In AGB stars $^{19}$F is
predicted to be produced in the convective He-rich intershell and then dredged-up to
the surface during the He-burning thermal pulses \citep{forest92}.
The reaction chain for F production in the He-burning environments of
AGB and W-R stars is: 
$^{14}$N($\alpha$,$\gamma$)$^{18}$F($\beta^+$)$^{18}$O(p,$\alpha$)$^{15}$N($\alpha$,$\gamma$)$^{19}$F.
Protons are provided by the $^{14}$N(n,p)$^{14}$C reaction with neutrons
liberated from $^{13}$C($\alpha$,n)$^{16}$O. The $^{14}$N and $^{13}$C nuclei
are the result of H-burning by CNO cycling, and the initial $^{13}$C stock
acts as a limiting factor for the $^{19}$F yield.

The Cunha et al.'s same fluorine abundance results for the Milky Way were later
discussed in \citet{renda2004} in light of a chemical evolution model for the Galaxy
that took into account all possible sources for fluorine production. This chemical
evolution model, however, cannot be applied to the peculiar globular cluster $\omega$\,Cen
and the low fluorine abundance results for $\omega$\,Cen remain puzzling. Renda et al.
suggest that both W-R and AGB stars are
significant sources of fluorine, and that is supported by the evidence of
enhanced F abundance from F\,{\sc v} and F\,{\sc vi} absorption lines in
the far-UV spectra of hot post-AGB PG\,1159 stars \citep{werner2005}.
Also, \citet{zhang2005} determine F abundances from the nebular emission
lines: [F\,{\sc ii}] line at 4789\AA\ and [F\,{\sc iv}] line at 4060\AA, for
a sample of Planetary nebulae (PNe). Fluorine is abundant in PNe, and this provides
evidence for the synthesis of fluorine in the AGB phase. No evidence of enhanced fluorine resulting 
from spallation in Type\,II SNe is suggested by the observations of F\,{\sc i} 
interstellar absorption line at 955\AA\ in two sight lines towards the Cep OB2
association using $Far$ $Ultraviolet$ $Spectroscopic$ $Explorer$ \citep{feder2005}.

Extreme helium (EHe) stars, see for \citet{pan2006} and the references therein, are
suggested to have gone through AGB phase in their earlier evolution, hence, fluorine
should be present in their atmospheres. Presence of fluorine in EHe's atmosphere
can serve as a test bed for fluorine production in AGB stars.
In this letter, we present fluorine abundances for the sample of cool EHe stars
from F\,{\sc i} lines.

\section{Observations}

High-resolution optical spectra of four cool EHes: FQ\,Aqr, LS\,IV-14$^{\circ}$\,109,
BD\,-1$^{\circ}$\,3438, and LS\,IV-1$^{\circ}$\,2, were obtained
at the W. J. McDonald Observatory's 2.7-m telescope with the
coud\'{e} cross-dispersed echelle spectrograph \citep{tull95} at a 2-pixel
resolving power (R =$\lambda/\Delta\lambda$) of 60,000.
The details of these observations are described in \citet{pan2001}.
The observations of the  cool EHe LSS\,3378, described
in \citet{panred2006}, were acquired at CTIO.

Finally, spectra of HD\,168476 $=$ PV\,Tel, the coolest member of the hot EHe stars,
were obtained with the Vainu Bappu Telescope (VBT) of the Indian Institute of
Astrophysics with a fiber-fed cross-dispersed echelle
spectrometer \citep{rao04,rao05b}. These spectra with a resolving power of about 30,000
were recorded on a 2048 $\times$ 4096 pixels CCD camera exclusively built for the
spectrometer. The signal-to-noise (S/N) per pixel of the final co-added spectrum,
from the exposures obtained on 2006 April 22 and 2006 May 15, is $\sim$40 at about 6902\AA.

In addition, McDonald spectra of HR\,2074, a normal A0 Ia-type supergiant, and
KS\,Per, a H-deficient hot primary of a spectroscopic binary, were
available for examination. The telluric absorption lines
from the spectra of the programme stars were removed interactively
using early-type rapidly rotating stars.
We have used the Image Reduction and
Analysis Facility (IRAF) software packages to reduce the spectra, and the task $telluric$ within
IRAF to remove the telluric absorption lines. In Figure 1, the telluric absorption lines from the
spectrum of the cool EHe LSS\,3378 are not removed just to show their presence.

\begin{figure}
\epsscale{1.00}
\plotone{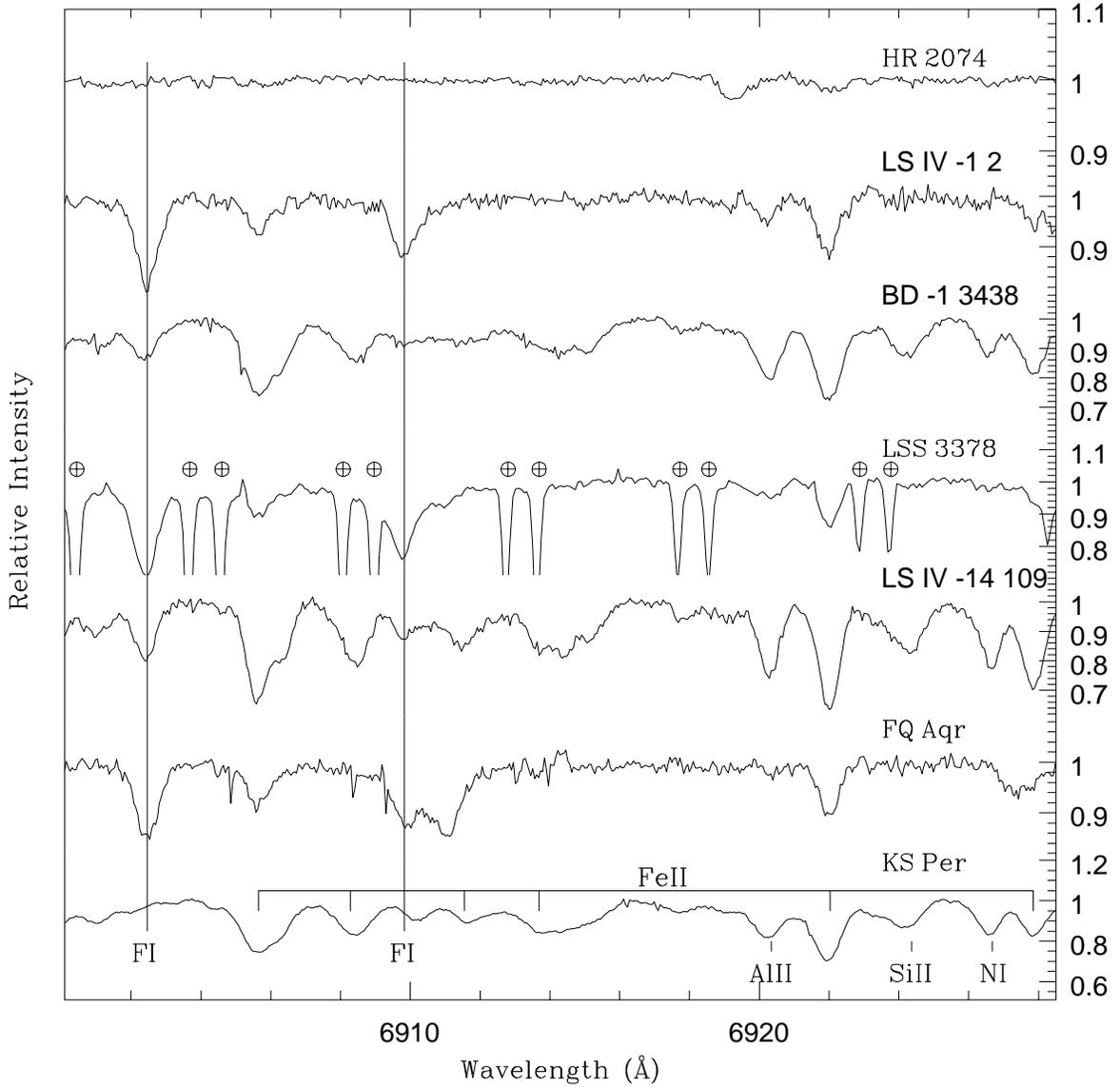}
\caption{A sample of F\,{\sc i} lines, that are indicated by vertical solid lines,
region of the spectra of cool EHe stars that are plotted with their effective
temperatures increasing from bottom to top. The positions of the key lines
are identified in this window from 6900 \AA\ to 6928 \AA. The telluric absorption lines
in the spectrum of LSS\,3378 are shown marked $\earth$. The spectra of KS\,Per and
HR\,2074 are also plotted. \label{fig1}}
\end{figure}

\section{F\,{\sc i} Lines}

The multiplets numbered 1, 2, 4, and 6 by \citet{moor72} are
potential contributors of F\,{\sc i} absorption lines to cool EHe stars' spectra.
Complete listings of the wavelengths for lines of these
multiplets from the 3$s$ -- 3$p$ transition array were compiled from
the NIST database\footnote{http://physics.nist.gov/PhysRefData/ASD/lines\_form.html}.
Measured $gf$-values were taken from \citet{musie99} who show that their
results are in fair agreement not only with the semiempirical calculations
of \citet{kur75} but also with earlier experimental determinations; the latter
on a relative scale.  Table 1 gives a partial list of these multiplets; lines of very
small $gf$-value are omitted on the ground that their F\,{\sc i}
contribution must be very small.

\clearpage
\begin{center}\small{Table 1} \\
F\,{\sc i} lines from $2s^22p^4$($^3$P)$3s$ -- $2s^22p^4$($^3$P)$3p$ transition array detected in the
spectra of the analysed stars. The F\,{\sc i} lines used in abundance
determination for all the analysed cool EHes are shown in bold\\
\begin{tabular}{ccccl}
\hline
\hline
Multiplet & $\lambda$ & $\chi$ & log $gf$ & Contributors \\
(No.)    & (\AA)     & (eV)   &          &              \\
\hline
$3s$ $^4$P -- $3p$ $^4$P$^{\circ}$ &{\bf 7482.72}&12.73& $-$0.66 & F\,{\sc i}, C\,{\sc i} $\lambda$7483.445\AA\ (red wing), Fe\,{\sc ii} $\lambda$7482.777\AA\ (weak)\\
(1) &7514.93&12.75& $-$0.96 & F\,{\sc i} \\
 &{\bf 7331.95}&12.70& $-$0.11 & F\,{\sc i}, N\,{\sc i} $\lambda$7332.055\AA, Fe\,{\sc ii} $\lambda$7332.115\AA\ (weak)\\
 &{\bf 7425.64}&12.73& $-$0.19 & F\,{\sc i}, Fe\,{\sc ii} $\lambda$7425.095\AA \\
 &7552.24&12.73& $-$0.34 & F\,{\sc i} \\
 &7573.41&12.75& $-$0.34 & F\,{\sc i} \\
 &       &     &         &            \\
$3s$ $^4$P -- $3p$ $^4$D$^{\circ}$ &{\bf 6856.02}&12.70& $+$0.44 & F\,{\sc i}, Fe\,{\sc ii} $\lambda$6855.646\AA \\
(2) &{\bf 6902.46}&12.73& $+$0.18 & F\,{\sc i} \\
 &{\bf 6909.82}&12.75& $-$0.23 & F\,{\sc i} \\
 &{\bf 6773.97}&12.70& $-$0.40 & F\,{\sc i}, Fe\,{\sc ii} $\lambda$6774.146\AA, Fe\,{\sc ii} $\lambda$6774.473\AA\ (red wing) \\
 &{\bf 6834.26}&12.73& $-$0.21 & F\,{\sc i} \\
 &6795.52&12.73& $-$1.09 & F\,{\sc i} \\
 &       &     &         &            \\
$3s$ $^2$P -- $3p$ $^2$D$^{\circ}$ &{\bf 7754.70}&12.98& $+$0.24 & F\,{\sc i}, Ti\,{\sc ii} $\lambda$7755.751\AA\ (red wing) \\
(4) &{\bf 7800.22}&13.03& $+$0.04 & F\,{\sc i}, Si\,{\sc i} $\lambda$7799.996\AA\ (weak), Fe\,{\sc ii} $\lambda$7801.235\AA\ (red wing) \\
 &             &     &         &       \\
$3s$ $^2$P -- $3p$ $^2$P$^{\circ}$ &{\bf 7037.45}&12.98& $+$0.10 & F\,{\sc i}, unknown blend (blue wing)\\
(6) &7127.88&13.03& $-$0.12 & F\,{\sc i} \\
 &6966.35&12.98& $-$1.01 & F\,{\sc i} \\
\hline
\end{tabular}
\end{center}
\clearpage

Eleven F\,{\sc i} lines were  identified as the principal or
leading contributor to a stellar line - see Table 1.
Figure 1 illustrates the window from 6900\AA\ to 6928\AA\ that represents a couple of F\,{\sc i} 
lines at 6902.46, and 6909.82\AA, where the cool EHe stars are ordered from top to bottom in order
of decreasing effective temperature.
All spectra are alligned to the rest wavelengths of the well known lines, that fall in
the wavelength regions. 
At the bottom, KS\,Per is added, which is known to be H-deficient \citep{wall67,pana2006}.
The spectrum of KS\,Per very much resembles the spectrum of
the cool EHe LS\,IV-14$^{\circ}$\,109, except for the C\,{\sc i} lines which are extremely
week or not present in the KS\,Per's spectrum \citep{wall67,pana2006}.
Singly ionized metal
lines (particularly Fe\,{\sc ii}) are identified as
contributor to a  stellar line at or about F\,{\sc i} lines in KS\,Per's spectrum.
Note the presence of almost no lines in the KS\,Per's spectrum at or about some contributing
F\,{\sc i} lines in cool EHes' spectra (Figure 1).
Only these F\,{\sc i} lines, if not blended by C\,{\sc i} lines, are identified as the sole major
contributor to a stellar line in a cool EHe's spectrum. The pattern to be contrasted
is, for example, the C\,{\sc i} and F\,{\sc i} lines show little variation in
these cool EHe stars when compared to the
metal lines (e.g., Fe\,{\sc ii}) that show a considerable star to star variation (Figure 1). Finally, the spectrum of HR\,2074, a normal A0 Ia-type supergiant
is shown at the top.
Lines of all elements expected and observed in the spectrum of a normal A0 Ia-type supergiant
were found. Lines of ionized metals of the iron group are plentiful.
These lines are much stronger when compared with those observed in HR\,2074, a notable
feature of the spectra of cool EHe stars and KS\,Per, and attributable to the
lower opacity in the atmosphere due to hydrogen deficiency.

For all F\,{\sc i} lines, an attempt was made to conduct a thorough search for blending lines.
Databases examined included the Kurucz's database\footnote{http://kurucz.harvard.edu},
tables of spectra of H, C, N, and O \citep{moor93}, selected tables of atomic spectra
of Si \citep{moore65,moore67}, the NIST database, and
the new Fe\,{\sc i} multiplet table \citep{nave94}. These provided not only wavelengths
of potential blends but also estimates of a line's $gf$-value.
The spectrum of supergiant HR\,2074 was also considered; F\,{\sc i} is not
represented in its spectrum.

The contributors to a stellar line at or about F\,{\sc i} lines are listed
in Table 1; the lines shown in bold are selected for abundance determination
for all the analysed cool EHes. The additional F\,{\sc i} lines listed in Table 1
are in the missing wavelength regions of the McDonald spectra, and, are only used
in abundance determination for the cool EHe LSS\,3378.

\section{Abundance analysis}

The fluorine abundances were determined from the
F\,{\sc i} lines listed in Table 1, by adopting the same procedure described
in \citet{pan2006}.
A model atmosphere is used with the Armagh LTE code SPECTRUM
\citep{jeff2001} to compute the equivalent width of a F\,{\sc i} line or
a synthetic spectrum for a selected spectral window. The synthetic spectrum was
convolved with a Gaussian profile, to account for the combined effects of the stellar
macroturbulence and the intrumental profile, before matching with the observed spectrum.
The fact that several F\,{\sc i} lines are blended with other lines
requires that the fluorine abundance be extracted by spectrum synthesis.
However, most of the F\,{\sc i} lines are only slightly blended in the iron poor
cool EHe stars: FQ\,Aqr, LSS\,3378, and LS\,IV-1$^{\circ}$\,2.

\begin{deluxetable}{llr}
\tabletypesize{\scriptsize}
\tablewidth{0pt}
\tablecolumns{3}
\setcounter{table} {1}
\tablecaption{The analysed EHe stars and the H-deficient binary KS\,Per, their stellar parameters,
and fluorine abundances}
\tablehead{
\colhead{Star} & \colhead{($T_{\rm eff}$, $\log g$, $\xi$)} & \colhead{log $\epsilon(\rm F)$}}
\startdata
BD+1$^{\circ}$\,4381 & (8750, 0.75, 8.0) & 6.45$\pm$0.12(11)\\
($=$FQ\,Aqr)&& \\
&&\\
V4732\,Sgr & (9500, 1.0, 6.0) & 6.52$\pm$0.19(11)\\
($=$LS\,IV-14$^{\circ}$\,109)&& \\
&&\\
CoD\,-48$^{\circ}$\,10153 & (10600, 0.4, 6.0) & 7.30$\pm$0.17(17) \\
($=$LSS\,3378)&& \\
&&\\
BD\,-1$^{\circ}$\,3438 & (11750, 2.3, 10.0) & 6.24$\pm$0.18(11) \\
($=$NO\,Ser)&& \\
&&\\
V2244\,Oph & (12750, 1.75, 10.0) & 7.15$\pm$0.14(11)\\
($=$LS\,IV-1$^{\circ}$\,2)&& \\
&&\\
HD\,168476 & (13750, 1.6, 10.0) & $<$7.20\\
($=$PV\,Tel)  & & \\
&&\\
HD\,30353  & (10500, 1.5, 10.0)\tablenotemark{a} & $<$4.40\\
($=$KS\,Per)  & & \\
\enddata
\tablenotetext{a}{from \citet{pana2006}}
\end{deluxetable}


\begin{figure}
\epsscale{1.00}
\plotone{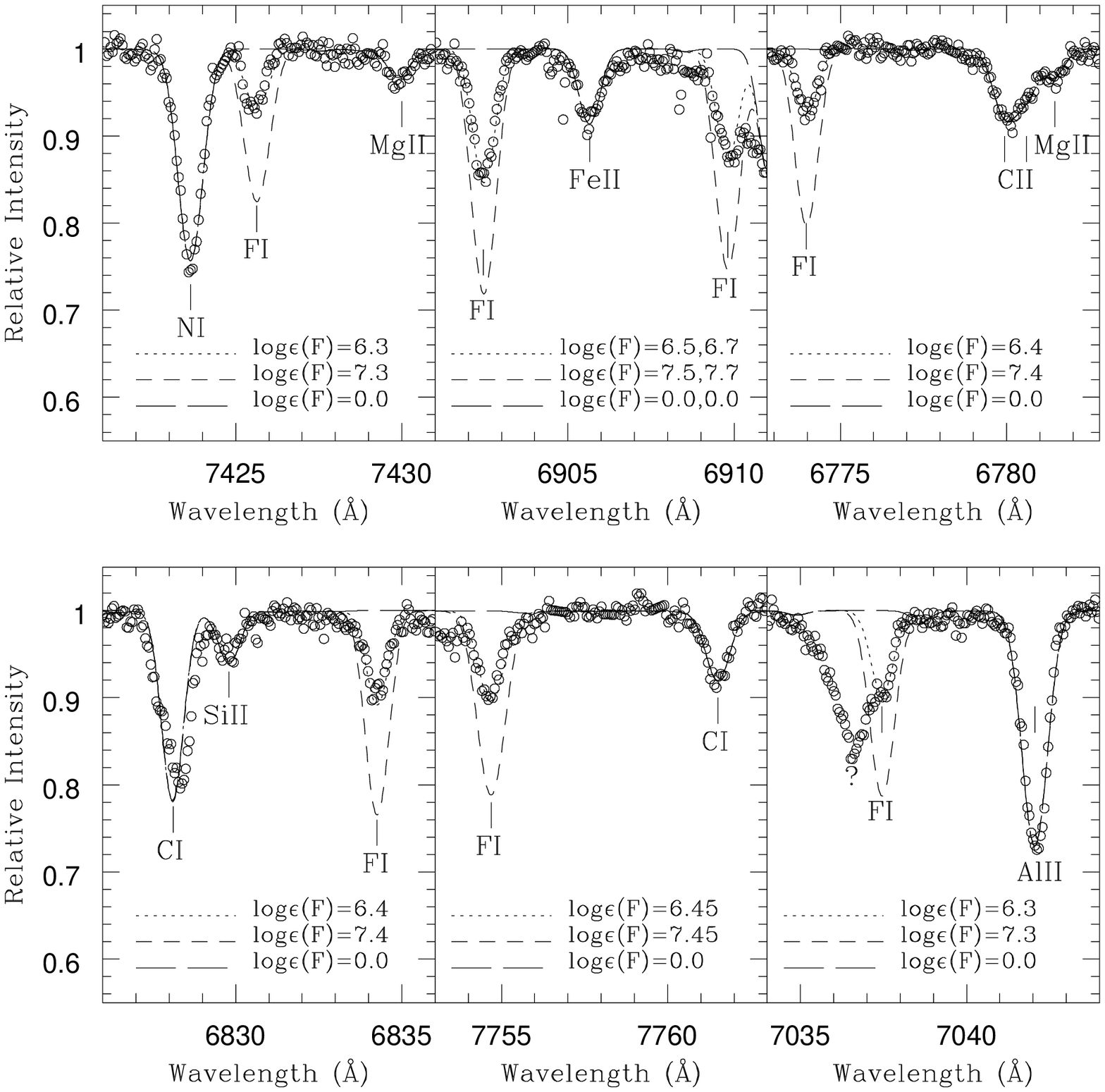}
\caption{Observed F\,{\sc i} wavelength regions (open circles) of the sample
cool EHe FQ\,Aqr with key lines marked. Synthetic spectra are shown for three fluorine 
abundaces, as shown on the figure. \label{fig2}}
\end{figure}

H-deficient model atmospheres have been computed
using the code STERNE \citep{jeff2001} for
the four stars with an effective temperature greater than 10,000 K (Table 2).
For  FQ\,Aqr with $T_{\rm eff} = 8750$ K, and LS\,IV-14$^{\circ}$\,109 with
$T_{\rm eff} = 9500$ K, we adopt the Uppsala model atmospheres \citep{asp97}.

The adopted stellar parameters: effective temperature ($T_{\rm eff}$),
surface gravity ($\log g$), microturbulent velocity ($\xi$) and
a carbon-to-helium abundance ratio C/He, are from \citet{pan2001}
for FQ\,Aqr, LS\,IV-14$^{\circ}$\,109, BD\,-1$^{\circ}$\,3438, and LS\,IV-1$^{\circ}$\,2,
from \citet{panred2006} for LSS\,3378, and from \citet{pan2006} for PV\,Tel.
Similarly, the elemental abundances affecting the blending lines were also taken
from the above references. The contribution of the blends were varied within limits
set by the abundances and the $gf$-values. The $gf$-values of the blending
lines Si\,{\sc i}, C\,{\sc i}, and rest of the lines (see Table 1), are from the
compilations by R. E. Luck (private communication), NIST database, and Kurucz's database,
respectively. An example of the best fitting profiles, for 7 of the studied
F\,{\sc i} wavelength regions, is illustrated in Figure 2 for the sample
cool EHe FQ\,Aqr.
The measured mean fluorine abundances (Table 2) are given as
log $\epsilon(\rm F)$, normalized such that log $\Sigma$$\mu_i \epsilon(i)$ = 12.15
where $\mu_i$ is the atomic weight of element $i$.
An upper limit to the F abundance for PV\,Tel is obtained by comparing
the synthetic and observed profiles of the F\,{\sc i} line at 6902.46\AA\ (Table 2).
An upper limit to the F abundance for KS\,Per, the H-deficient binary that shows
pure hydrogen burnt CN- and ON-cycled material, is also estimated from the
nondetection of the F\,{\sc i} line at 6902.46\AA\ (Figure 1, Table 2).
The line-to-line abundance scatter, and the number of lines used given within brackets, are in
column 3 of Table 2. The effective error in F abundances due to uncertainty in the
adopted stellar parameters, that are typically: $\Delta$$T_{\rm eff}$ = $\pm$500 K,
$\Delta$$\log g$ = $\pm$0.25 cgs and $\Delta$$\xi$ = $\pm$2 km s$^{-1}$, is about 0.2 to 0.3 dex.
Note that the derived fluorine abundances in LTE represent only the first step defining the
absolute F abundance in these stars; more reliable values should, in principle, come from
full non-LTE calculations.

\section{Discussion}

A good number of F\,{\sc i} lines were identified for the first time in the optical 
spectra of the sample of cool EHe stars. An abundance analyses of these spectra
provide the first abundances of fluorine for cool EHe stars. The enhanced F abundances
(see Figure 4) in cool EHes are evidence of nucleosynthesis in their earlier
evolution. The abundances for the other elements presented in Figure 4 are adopted
from \citet{pan2001} for FQ\,Aqr, LS\,IV-14$^{\circ}$\,109, BD\,-1$^{\circ}$\,3438, and 
LS\,IV-1$^{\circ}$\,2, from \citet{panred2006} for LSS\,3378, and from \citet{wall67} and
\citet{pana2006} for KS\,Per. The F, C, and O abundances in cool EHes are independent of the star's
initial metallicity (Fe abundance); the N abundances, however, show a clear trend with Fe
(see Figure 4). These trends clearly suggest that F, C, and O abundances in cool EHes
are associated with He-burnt material, and the N abundances are the result of
H-burning via CNO cycling; the almost complete conversion of the initial C, N, and O
to N is shown in the plot [N] vs [Fe] of Figure 4. During the He-burning thermal pulses $^{19}$F
is produced, and the initial $^{13}$C supply then acts as a limiting factor for the
$^{19}$F yield. To account for the enhanced F abundances, in cool EHes, with no trend 
with Fe, requires a primary $^{13}$C supply instead of the secondary $^{13}$C supply
from the CNO cycle ashes. Note the [F] vs [H] trend of Figure 4.

\begin{figure}
\epsscale{1.00}
\plotone{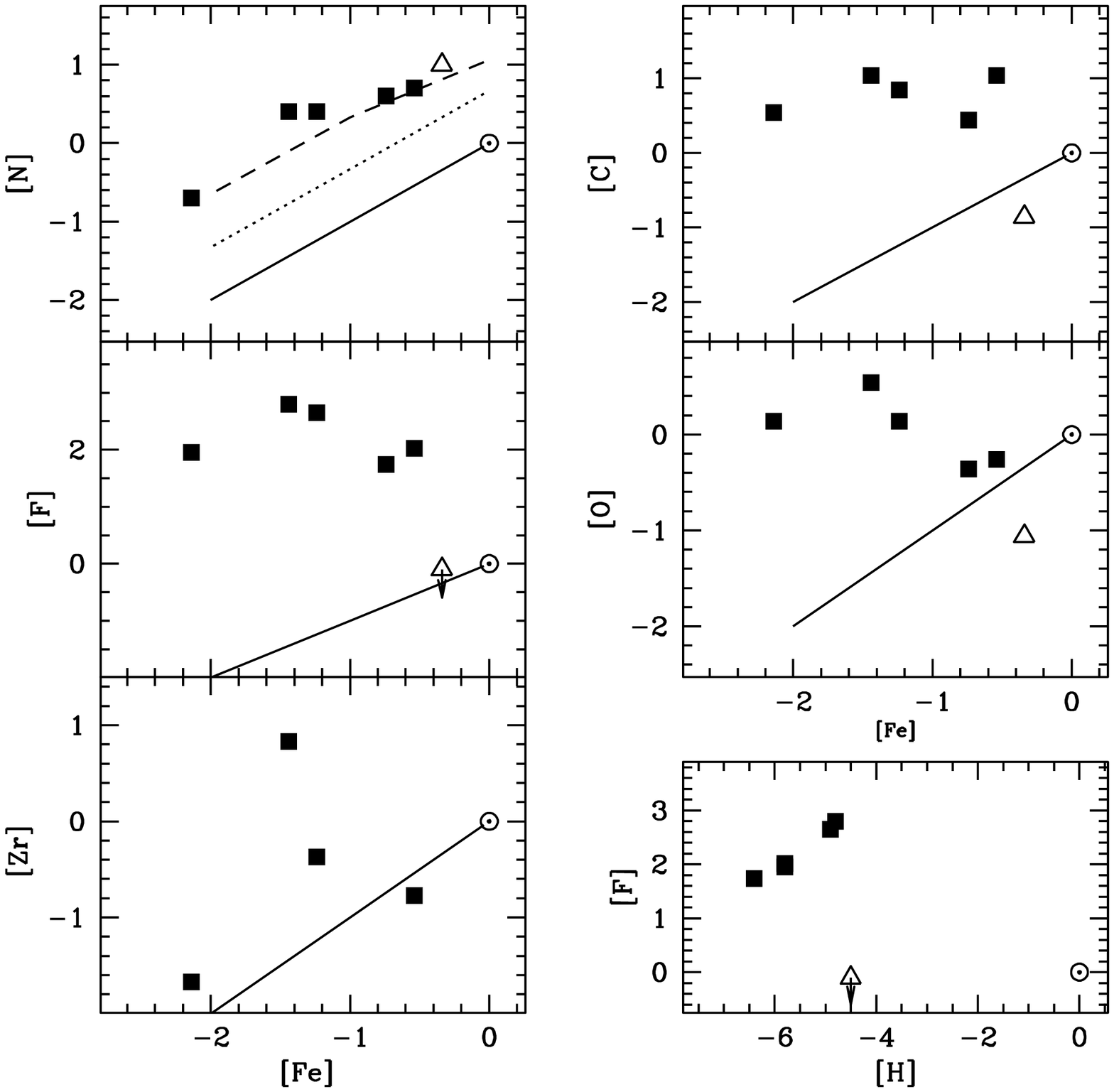}
\caption{[N], [C], [F], [O], and [Zr] vs [Fe], and also [F] vs [H]. The sample of
cool EHes is represented by filled squares; the abundances are from the references
given in Section 4. KS\,Per is represented by open triangles; the abundances are from 
\citet{wall67}, and \citet{pana2006} in preparation.
$\sun$ represents Sun; the adopted solar abundances are from Table 1 of \citet{lod2003}.
The dotted line represents conversion of the initial sum of C and N to N. The dashed line
represents the locus of the sum of initail C, N, and O converted to N. [X] = [Fe] are 
denoted by the solid lines where X represents N, C, F, O, and Zr. \label{fig3}}
\end{figure}

The F abundances in H-deficient PG\,1159 stars show a strong variation from star to star,
ranging from solar to 250 times solar \citep{werner2005}. The sample cool EHes show F
enhancements about 100 times solar. Note that the models of \citet{lug2004}, for thermally
pulsating AGB stars, predict that the F enrichment in the intershell is a strong function
of the initial stellar mass; for solar metallicity models, log $\epsilon(\rm F)$, which is 
about 100 times solar, is maximal at M = 3$M_\odot$. 
Lugaro et al. conclude that their F intershell abundances are not high
enough to explain the enhanced F abundances which were measured in the AGB
stars from the Jorissen et al. sample.
The surface composition of these 
H-deficient stars: PG\,1159 and EHe, could be a result of final He-shell flash in a single
post-AGB star (FF scenario), or a merger of two white dwarfs (DD scenario). Although the
FF scenario accounts for PG\,1159 stars, present theoretical calculations \citep{herw99} imply
higher C/He and O/He ratios than are observed in EHe stars. However, DD scenario can
account for the observed surface abundances of most of the elements in EHes \citep{pan2006}.

Enrichment of $s$-process elements, and also fluorine, is not expected for
the EHes resulting from a merger of a He with a C-O white dwarf (DD scenario) as discussed
by \citet{pan2004,pan2006}. However, synthesis of fluorine and $s$-process elements
may occur during the merger and needs to be explored. The $s$-process elements
are enriched in the cool EHe LSS\,3378 (see for example, [Zr] vs [Fe] of Figure 4).
The neutron source in EHes showing enriched $s$-process elements is not known.
$^{13}$C($\alpha$,n)$^{16}$O and $^{22}$Ne($\alpha$,n)$^{25}$Mg are the two possible sources.
If the latter reaction dominates, then F is unlikely to survive in the He-rich intershell, as 
the reaction rate of $^{19}$F($\alpha$,n)$^{22}$Ne is much higher than that of
$^{22}$Ne($\alpha$,n)$^{25}$Mg \citep{jori92}.
Since $^{19}$F is overabundant in cool EHes, [an upper limit to the F abundance for the coolest 
hot EHe PV\,Tel is also derived (see Table 2)], the reaction $^{22}$Ne($\alpha$,n)$^{25}$Mg is 
unlikely to be the neutron source in EHe stars. Non-LTE calculations
should be performed for the key elements to further improve the chemical analysis.

\acknowledgments

GP is greatful to the referee for the comments and suggestions.
GP thanks David Lambert and Kameswara Rao for the McDonald spectra, and Eswar Reddy
and Swara Ravindranath for commenting on a draft of this letter.


\clearpage

\begin{thebibliography}{}
\bibitem[Asplund et al.(1997)]{asp97} Asplund, M., Gustafsson, B., Kiselman, D.,
\& Eriksson, K. 1997, \aap, 318, 521
\bibitem[Cunha et al.(2003)]{cunha2003} Cunha, K., Smith, V. V., Lambert, D. L.,
\& Hinkle, K. H. 2003, \aj, 126, 1305
\bibitem[Federman et al.(2005)]{feder2005} Federman, S. R., Sheffer, Y., Lambert, D. L.,
\& Smith, V. V. 2005, \apj, 619, 884
\bibitem[Forestini et al.(1992)]{forest92} Forestini, M., Goriely, S., Jorrisen, A.,
\& Arnould, M. 1992, \aap, 261, 157
\bibitem[Hall \& Noyes(1969)]{hall69} Hall, D. N. B., \& Noyes, R. W. 1969,
\apjl, 4, L143
\bibitem[Herwig et al.(1999)]{herw99} Herwig, F., Bl\"{o}cker, T., Langer, N., \&
Driebe, T. 1999, \aap, 349, L5
\bibitem[Jeffery, Woolf \& Pollacco(2001)]{jeff2001} Jeffery, C. S., Woolf, V. M.,
\& Pollacco, D. L. 2001, \aap, 376, 497
\bibitem[Jorissen, Smith \& Lambert(1992)]{jori92} Jorissen, A., Smith, V. V., \&
Lambert, D. L. 1992, \aap, 261, 164
\bibitem[Kurucz \& Peytremann(1975)]{kur75} Kurucz, R. L., Peytremann, E. 1975,
SAO -- Special Report No. 362
\bibitem[Lodders(2003)]{lod2003} Lodders, K. 2003, \apj, 591, 1220
\bibitem[Maynet \& Arnould(2000)]{may2000} Maynet, G., \& Arnould, M. 2000,
\aap, 355, 176
\bibitem[Lugaro, Ugalde \& Karakas(2004)]{lug2004} Lugaro, M., Ugalde, C., \&
Karakas, A., I. 2004, \apjl, 615, L934
\bibitem[Moore(1965)]{moore65} Moore, Ch. E. 1965, Selected Tables of Atomic Spectra,
NSRDS--NBS 3, Section 2, Washington
\bibitem[Moore(1967)]{moore67} Moore, Ch. E. 1967, Selected Tables of Atomic Spectra, 
NSRDS--NBS 3, Section 1, Washington
\bibitem[Moore(1972)]{moor72} Moore, Ch. E. 1972, A Multiplet Table of Astrophysical
Interest, NSRDS--NBS, Washington
\bibitem[Moore(1993)]{moor93} Moore, Ch. E. 1993, Tables of Spectra of Hydrogen, Carbon, 
Nitrogen, and Oxygen Atoms and Ions, Editor: Jean W. Gallagher, CRC Series in Evaluated Data
in Atomic Physics, CRC press 
\bibitem[Musielok et al.(1999)]{musie99} Musielok, J., Pawelec, E., Griesmann, U., \&
Wiese, W. L. 1999, \pra, 60, 947
\bibitem[Nave et al.(1994)]{nave94} Nave, G., Johansson, S., Learner, R. C. M., 
Thorne, A. P., \& Brault, J. W. 1994, \apjs, 94, 221
\bibitem[Pandey et al.(2001)]{pan2001} Pandey, G., Rao, N. K., Lambert, D. L.,
Jeffery, C. S., \& Asplund, M. 2001, \mnras, 324, 937
\bibitem[Pandey et al.(2004)]{pan2004} Pandey, G., Lambert, D. L., Rao, N. K.,
\& Jeffery, C. S. 2004, \apjl, 602, L113
\bibitem[Pandey et al.(2006a)]{pan2006} Pandey, G., Lambert, D. L.,
Jeffery, C. S., \& Rao, N. K. 2006a, \apj, 638, 454
\bibitem[Pandey \& Reddy(2006)]{panred2006} Pandey, G., \& Reddy, B. E. 2006,
\mnras, 369, 1677
\bibitem[Pandey et al.(2006b)]{pana2006} Pandey, G., et al. 2006b in preparation
\bibitem[Rao et al.(2004)]{rao04} Rao, N. K., Sriram, S., Gabriel, F., Prasad,
B. R., Samson, J. P. A., Jayakumar, K., Srinivasan, R., Mahesh, P. K.,
\& Giridhar, S. 2004, Asian Journal of Physics, 13, 267
\bibitem[Rao et al.(2005)]{rao05b} Rao, N. K., Sriram, S., Jayakumar, K.,
\& Gabriel, F. 2005, JAA, 26, 331
\bibitem[Renda et al.(2004)]{renda2004} Renda, A., et al. 2004, \mnras, 354, 575
\bibitem[Tull et al.(1995)]{tull95} Tull, R. G., MacQueen P. J., Sneden, C., \&
Lambert, D. L. 1995, \pasp, 107, 251
\bibitem[Wallerstein, Greene \& Tomley(1967)]{wall67} Wallerstein, G., Greene, T. F.,
\& Tomley, L. J. 1967, \apj, 150, 245
\bibitem[Werner, Rauch \& Kruk(2005)]{werner2005} Werner, K., Rauch, T., \&
Kruk, J. W. 2005, \aap, 433, 641
\bibitem[Woosley \& Weaver(1995)]{woos95} Woosley, S. E., \& Weaver, T. A. 1995,
\apjs, 101, 181
\bibitem[Zhang \& Liu(2005)]{zhang2005} Zhang, Y., \& Liu, X.-W. 2005, \apjl, 631, L61


\end{thebibliography}
\end{document}